# An investigation into the turbulence induced by moving ice floes

Trygve K. Løken[1], Aleksey Marchenko[2], Thea J. Ellevold[1], Jean Rabault[3,1], Atle Jensen[1]
[1] University of Oslo, Oslo, Norway
[2] The University Centre in Svalbard, Longyearbyen, Norway
[3] Norwegian Meteorological Institute, Oslo, Norway

**ABSTRACT**

Several phenomena are known to attenuate waves in the marginal ice zone, such as wave scattering due to ice floes, viscous dissipation inside the ice and energy dissipation caused by ice floe interactions. Our aim of this study is to investigate colliding ice floes and to quantify the Turbulent Kinetic Energy (TKE) dissipation in the surrounding water through direct observations. The field work was carried out in Van Mijen Fjord on Svalbard, where an artificial 3x4 m$^2$ ice floe was sawed out in the 1 m thick ice. Wave motion was simulated by pulling the ice floe back and forth in an oscillatory manner with two electrical winches in a 4x6 m$^2$ pool. Ice floe motion was measured with a range meter, and the water turbulence was measured with acoustic velocimeters. TKE frequency spectra were found to contain an inertial subrange where energy was cascading at a rate proportional to $f^{-5/3}$. From the spectra estimated at several vertical positions, the TKE dissipation rate was found to decrease exponentially with depth. The total TKE dissipation rate was estimated by assuming that turbulence was induced over an area spanning over the ice floe width and the pool length. The preliminary energy budget suggests that approximately 60% of the input power from the winches was dissipated in turbulence, which experimentally confirms that energy dissipation by induced turbulent water motion is an important mechanism for colliding ice floe fields.

**INTRODUCTION**

A decline in the Arctic ice cover has been observed over the past decades, which has allowed for more human activities in the region, such as shipping, tourism and exploitation of natural resources (Smith and Stephenson, 2013; Feltham, 2015). The retreating ice cover leads to larger areas of open water in the Arctic where more energetic waves are generated due to the increased fetch, which in turn enhance ice break up processes (Thomson and Rogers, 2014). Higher waves can be hazardous to ships and navigation. Reliable wave forecast models are necessary to ensure safe human operations at sea, also in the Marginal Ice Zone (MIZ), which is the transition between the land fast ice or dense pack ice and the open ocean. The MIZ consists of various concentrations and types of ice, such as grease ice and ice floes.

Previous experiments have shown that waves are exponentially damped in the MIZ (Squire and Moore, 1980; Wadhams et al., 1988). Several phenomena are known to attenuate waves in the MIZ, such as wave scattering due to ice floes, viscous dissipation inside or in the boundary layer under the ice and energy dissipation caused by ice floe interactions (Squire et al., 1995). Rabault et al. (2019) showed from wave tank experiments that the dynamics of ice can generate turbulence that inject eddy viscosity in the water, which lead to enhanced energy dissipation. However, there are few in situ observations of the water kinematics around interacting ice floes, and there is uncertainty associated with the dominating source of wave energy dissipation. Increased knowledge about these processes, and hence atmosphere-wave-ice-ocean energy transfer, is necessary to improve sea ice dynamics models used for wave forecasts and climate modeling.

This study presents direct observations of the water kinematics around a moving ice floe. The floe was driven back and forth in an oscillatory manner to simulate wave induced motion in an ice floe field. As the floe moved relative to the surrounding water and collided with the walls, turbulence was induced around the sharp edges. The Turbulent Kinetic Energy (TKE) was measured with an Acoustic Doppler Current Profiler (ADCP) and an Acoustic Doppler Velocimeter (ADV). The motion of, and force applied on the ice floe was monitored with various sensors. This enabled us to estimate the ratio of input energy to the system over dissipated energy due to turbulence.

Fluid flow was also measured with an optical method where bubbles tracers were filmed with a remotely operated vehicle maneuvered below the ice. These data are currently being analyzed, in a similar fashion to Løken et al. (2021), who presented and validated the methodology.

## DATA AND METHODS

Ice floe towing experiments were carried out in the Van Mijen Fjord near Sveagruva on Svalbard in March 2020. The inner part of the fjord near Svea Bay was covered with 1.0 m thick ice. Figure 1 shows the location and preparation of the site, which was selected near the Svea harbor for practical purposes (seen in the background of Fig. 1b). An artificial ice floe was made by sawing out a 6x4 m$^2$ pool in the ice with a 4x3 m$^2$ floe left in the center. A 10x6 m$^2$ inflatable tent was placed over the pool to provide shelter for the researchers and the equipment.

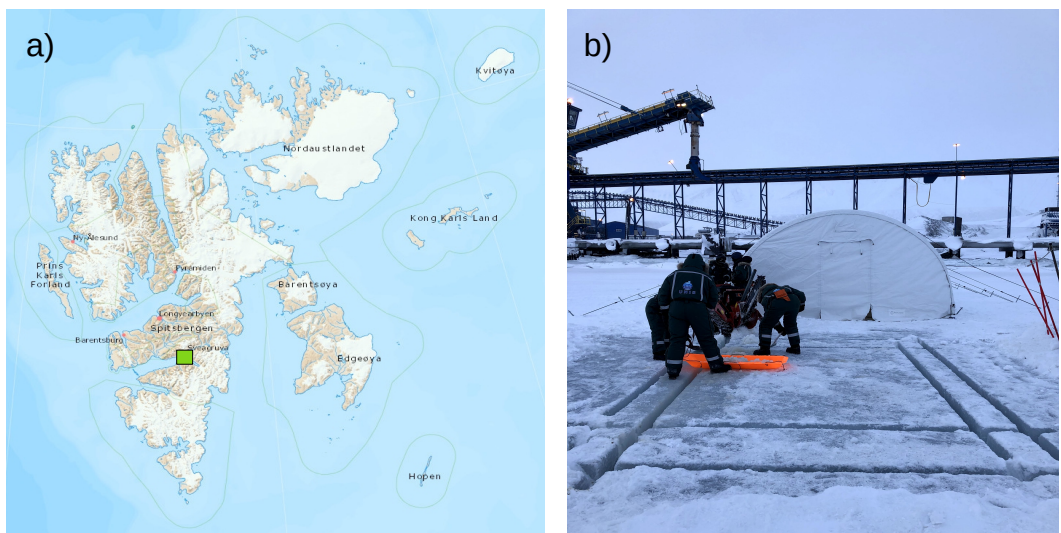

Figure 1. Location of the experiments. a) Map of the Svalbard Archipelago with the site indicated (TopoSvalbard, 2021). b) Photo of the site preparation where the pool and ice floe are made.

### Experimental setup

Two electrical winches were used to pull the ice floe back and forth in the pool length direction (axial direction) to simulate wave motion. The winches were mounted on steel frames that were fixed to the ice with ice screws, one on each side of the pool short ends. Each winch was powered by a large car battery (~80 Ah), which was continuously charged from a gasoline generator. The winch hooks were connected to a polyester-silk rope that was tied around a wooden frame, which was mounted to the ice floe with ice screws to distribute the pulling load. Figure 2a shows the pulling system. Two persons actuated the winches manually and tried to make the ice floe oscillate smoothly back and forth.

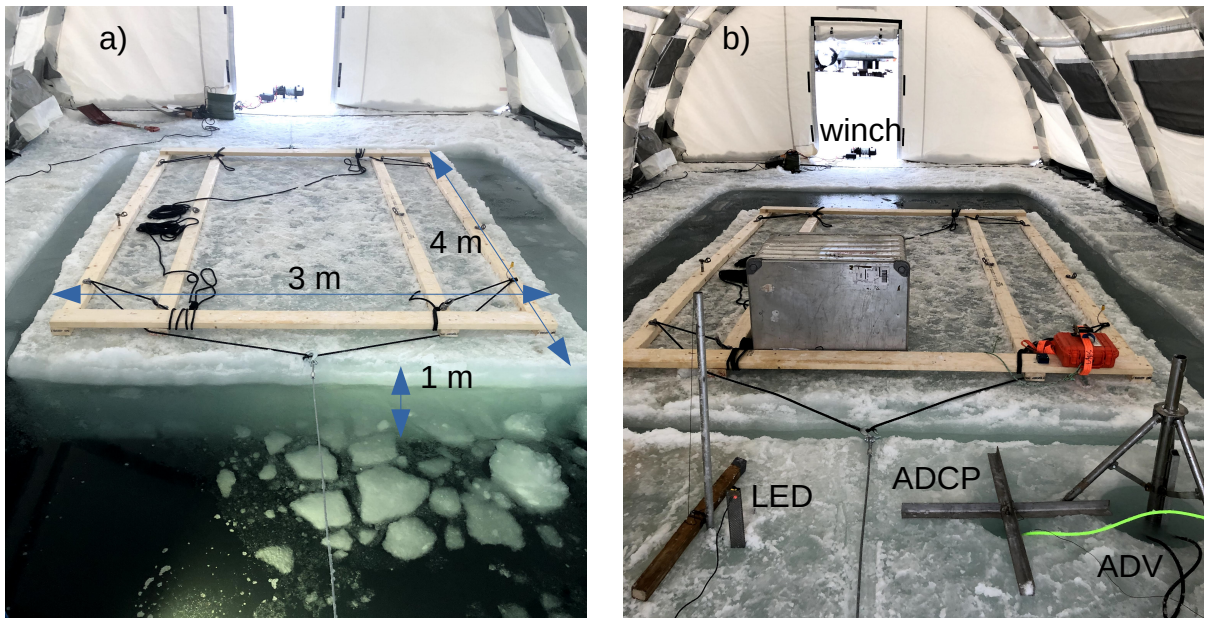

Figure 2. Experimental setup. a) Ice floe with pulling system. b) Instrumentation.

Figure 2b shows the laboratory instrumentation. Ice floe position in the axial direction was measured with an Evo60 LED range meter. The LED beam was pointing towards a large box that was placed on the ice floe. The water velocity around the ice floe was measured with a Nortek Signature 1000 kHz ADCP and a SonTek 5 MHz ADV, both placed 0.5 m from the pool. The ADCP was oriented downwards with one beam in the vertical direction and four beams slanted 25° off the vertical, two in the axial and two in the transverse direction. In some runs, the ADCP was placed 0.25 m from the pool, and in the last run, it was placed in the ice floe center. The ADCP range was 2 m, starting from 13 cm below the bottom ice, and the cell size was either 2 or 5 cm, corresponding to 95 and 39 cells, respectively. The ADV measurement volume was located 58 cm below the bottom ice. The two acoustic velocimeters were arranged such that one of the transverse ADCP beams intersected the ADV measurement volume, which allowed for a direct comparison of the time series. A PCM BD-ST-620 load cell (max load 2 kN) was used in one run to measure the tension on one of the winch wires. It was attached between the winch hook and the polyester-silk rope. The different experiments (E1-6) are summarized in Table 1.

Table 1. Experimental details.

| Exp. | Date | ADCP position | ADCP cells | Load cell | Cycles | Cycles ADV |
|------|------|---------------|------------|-----------|--------|------------|
| E1 | 7 | 0.5 m | 95 | - | 15 | 15 |
| E2 | 8 | 0.5 m | 39 | - | 14 | 14 |
| E3 | 9 | 0.5 m | 95 | yes | 11 | 5 |
| E4 | 10 | 0.25 m | 39 | - | 28 | 20 |
| E5 | 10 | 0.25 m | 39 | - | 7 | 5 |
| E6 | 11 | floe center | 39 | - | 8 | - |

**Data processing**

The range meter sample frequency was approximately 125 Hz. The raw displacement signal was smoothed with a moving average, where each mean was calculated over a sliding window of 200 data points. Ice floe velocity in the axial direction was obtained from the smoothed displacement with a central difference scheme, and then smoothed afterwards with the same method. The load cell sample frequency was 5 kHz and its signal was smoothed with a moving average with 500 data points window size. The towing power applied on the ice floe was calculated as towing force times floe velocity, and the work applied on the floe in the process of towing in one direction was found by numerically integrating the towing power with respect to time over one half cycle.

ADCP and ADV data were sampled at 8 and 10 Hz, respectively. Spikes in the time series, defined as 3 standard deviations (σ) off the moving mean (10 data point windows), were cut at ±3σ off the moving mean for spectral analysis, which require continuous time series (Nystrom et al., 2007). Turbulence properties were found from the fluctuating velocity components. Power Spectral Densities (PSDs) of the fluctuating velocities in the vertical direction $w$, also known as Turbulent Kinetic Energy (TKE) spectra, were calculated from the time series. The time series were subdivided into 50 s segments with 50% overlap and the segments were Fourier transformed. The spectral estimates were ensemble averaged to decrease statistical uncertainty according to the Welch method, and a Hanning window was applied to each segment to reduce spectral leakage (Earle, 1996). The resulting PSDs had approximately 6-28 degrees of freedom, depending on the number of cycles in the experiments (5-20).

The TKE spectra represents the energy cascading from the large turbulent eddies driven by the mean flow via the inertial subrange to the micro-scale where turbulence is dissipated into heat. The structure size is represented by the frequency $f$ at which the eddies rotate. At the inertial subrange, the TKE spectra are proportional to $f^{-5/3}$, and depends only on the TKE dissipation rate $\varepsilon$ and $f$ through the relation

$$PSD(f) = C_K \varepsilon^{2/3} f^{-5/3} \frac{w_{rms}^{2/3}}{2\pi}, \qquad (1)$$

where $C_K = 0.53$ is the universal Kolmogorov constant (Sreenivasan, 1995) and $w_{rms}$ is the root mean square of the fluctuating vertical velocity, which can be used in the case when the mean flow advecting past the instruments is nearly zero (Tennekes, 1975), like it was around the oscillating ice floe. From Eq. 1, $\varepsilon$ can be found by calculating the mean compensated TKE spectra $PSD(f)f^{5/3}$ averaged over the range of frequencies where the spectra are proportional to $f^{-5/3}$, i.e. in the inertial subrange.

**Instrument synchronization**

All the instruments sampled continuously and were manually started before the experiments were initiated, except the ADV which was configured with a fixed measurement interval of 11.67 min with 10 min continuous sampling. Therefore, the ADV did not sample all the cycles in a run if the instrument down-time coincided with the experiment. The total amount of cycles and cycles sampled by the ADV are listed in Table 1. For the comparison of turbulent flow properties obtained from the two acoustic velocimeters, time series containing the same number of cycles were used in the analysis, even though the ADCP sampled during the entire experiments.

The acoustic velocimeters were not synchronized in time. Therefore, this was performed in the post-processing, in order to compare time series of the same velocity component from the two instruments which coincided in space. The ADV time series were synchronized with the ADCP time series with a cross-correlation optimization method. This was achieved by re-sampling the two signals to a common sampling frequency $f = 80$ Hz. Then, the cross-correlation function $R(n\Delta t)$, where $n$ is an integer and $\Delta t = 1/f$, of the ADV and the lagged ADCP signal was found. The $n$ that yields a maximum in *R* indicates the time lag *nΔt* between the two signals where they are the most similar. The ADV signal was then shifted with the optimal time lag. An example of the synchronized raw data time series from E3 is presented in Fig. 3.

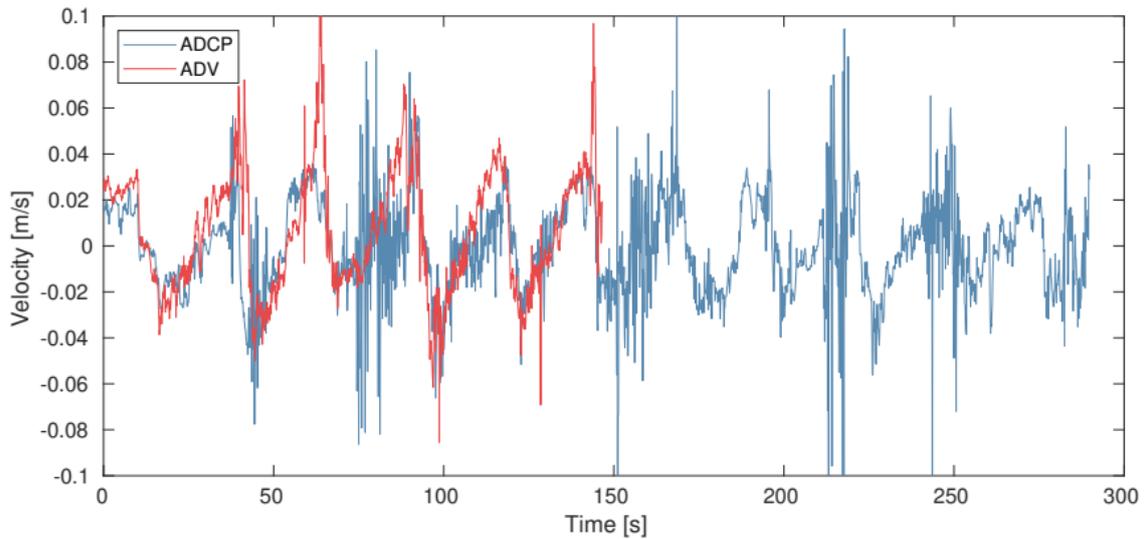

Figure 3. Synchronized ADV and ADCP raw data from E3. The ADV time series is shorter due to the pre-configured measurement scheme, as seen in Table 1. The velocity component is along the ADCP beam pointing in the transverse direction towards the ADV.

Time synchronization in the post-processing was also necessary with the range meter and load cell data. The signals were re-sampled to a common frequency of 1 kHz before the cross-correlation optimization method was applied to the displacement and load signals.

**RESULTS**

Figure 4 presents ice floe displacement (upper panel) and velocity in the axial direction, towing load and power (lower panel) during E3 where the ice floe undergoes 11 full periods.

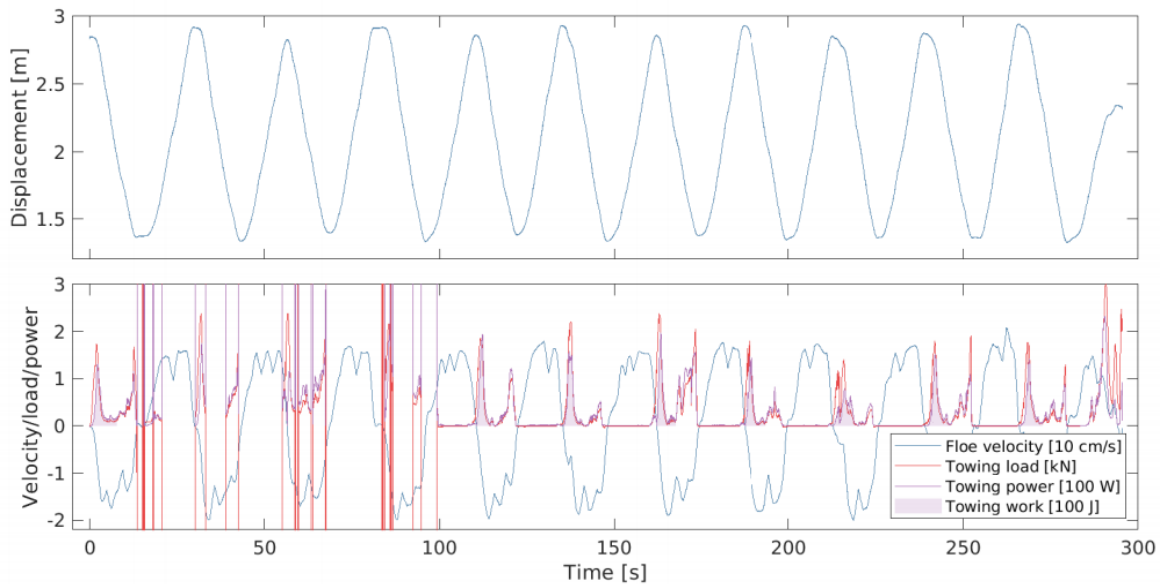

Figure 4. Synchronized range meter and load cell data from E3. Upper panel: Floe axial displacement. Lower panel: Floe axial velocity (blue), towing load (red), power (purple) and work (shaded purple). Some dropouts occurred in the load cell data in cycle 2-4.

The floe was displaced approximately 1.7 m and the maximum towing velocity was quite constant and about 0.15 m/s in each direction. The oscillating period was approximately 26 s. All the experiments (E1-6) were relatively consistent in terms of maximum towing velocity and oscillating period.

The shaded areas under the power curve in Fig. 4 indicates the towing work performed on the ice floe by the winch, which occurred when the floe was towed towards the range meter (i.e. decreasing displacement and negative velocity). In each cycle, there was typically one large peak in towing power due to the acceleration of the ice floe, succeeded by a smaller peak. The second peak occurred if the winch on the opposite side of the ice floe started to pull the floe in the reversed direction before the winch on the load cell side started to release wire. This situation mainly caused tension to the system, but did not transfer kinetic energy to the ice floe and is therefore neglected in the calculation of the work. When cycles 1 and 5-11 are considered (cycle 2-4 contained some dropouts), the average work applied to tow the ice floe in one direction was 263 J. One half cycle lasted on average 13.3 s, which means that the average power transfer from the winch to the ice floe was approximately 19.8 W. It is assumed that the winch power was constant in E1-6 due to the consistency in the runs.

TKE spectra were calculated from the fluctuating vertical velocity component from all the ADCP cells and from the ADV with the Welch method described under "Data processing". Figures 5a-b shows the spectra from E4 and E6, respectively, where only 10 ADCP cells evenly distributed over the 2 m deep profile are presented. The thicker orange spectrum in Fig. 5a is produced from the ADV data. The gray shaded region illustrates the range of frequencies over which the compensated spectra are averaged in order to estimate $\varepsilon$, i.e. where a slope proportional to $f^{-5/3}$ is expected. ADCP data quality typically deteriorate as the distance from the instrument increase. In E4, the spectra from the cells below -1.2 m flattened out towards the higher frequencies, which illustrates that the instrument noise level exceeded the TKE level. These data are not physical, hence not used to estimate $\varepsilon$. However, all the spectra in E6, when the ADCP was placed in the ice floe center, are approximately proportional to the $f^{-5/3}$ slope within the shaded region and are therefore used to estimate $\varepsilon$ from Eq. 1 along the entire profile.

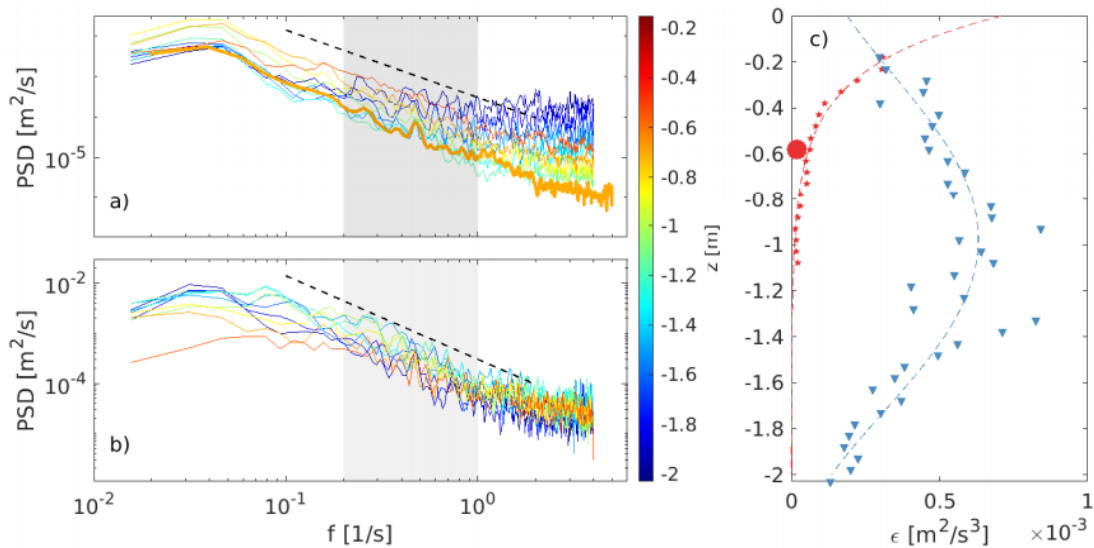

Figure 5. TKE spectra obtained with the ADCP from various depths during E4 (a) and E6 (b). The black dashed lines show the theoretical $f^{-5/3}$ slope. The thick orange graph in a) is obtained from the ADV spectrum. The shaded regions show the range of frequencies over which the compensated spectra are averaged to estimate $\varepsilon$. c) Estimated $\varepsilon$ from E4 (red) and E6 (blue) at each ADCP cell and from the ADV in E4 (red circle). The dashed lines are curve fits to the data.

Figure 5c shows the TKE dissipation rate $\varepsilon$ along the vertical profile of the ADCP in E4 (red) and E6 (blue). The red circle shows the estimated $\varepsilon$ from the ADV at the depth of its measurement volume. Estimated $\varepsilon$ decrease exponentially with depth in E4 when the ADCP was located next to the pool and increase up to a maximum at -1 m before it decreases with depth in E6 when the instrument was placed on the ice floe. Nonlinear regression by means of iterative least squares have been applied to fit an exponential and a fourth order polynomial function to the estimated $\varepsilon$ in E4 and E6, respectively, shown with dashed lines in Fig. 5c.

## DISCUSSION

Data from the two acoustic velocimeters agree quite well. In the time series presented in Fig. 3, both instruments resolve the same large turbulent structures, but there are some discrepancies, especially on the smaller scales, probably due to the approximately two orders of magnitude larger ADCP measurement volume. In the TKE spectra shown in Fig. 5, the two instruments produced similar slopes in E4, although the ADV energy level is slightly lower than the corresponding ADCP cell. The same observation can be made from the TKE dissipation rate. This deviation can be explained by the fact that the ADCP was placed closer to the pool than the ADV, where the TKE level is expected to be higher. Ideally, the ADV should have been mounted at the same distance to the pool as the ADCP, but this was not possible with the tripod that was available for the ADV. Also, for a better comparison of the spectral data along the profile, the ADV should have been deployed at different depths. However, time limitations at the site did not allow for this.

As the ice floe was displaced back and forth, an oscillating flow was generated in the pool. The shear flow around the solid boundaries of the ice floe and the pool transferred energy to large turbulent structures with the same time scale as the towing motion, which can be seen from the peak frequency in the TKE spectra (26 s periods). As the ice floe approached the pool end, turbulent downward jets were forming in the closing gap. The two TKE dissipation rate profiles presented in Fig. 5c are very different, probably due to the location of the ADCP. When the instrument was placed next to the pool, it primarily recorded the jets, which explains the decay with depth that was also observed in E1-E5 (not shown). When the ADCP was placed in the ice floe center, the dominant source of turbulent energy was likely from the sharp edges of the moving floe. Here, TKE would decay with depth beyond the turbulent boundary level, but it first increased with depth immediately below the ice.

The total TKE dissipation rate caused by the ice floe motion can be estimated from the product of $\varepsilon$, the total water volume where the dissipation occurred and the water density. Ideally, $\varepsilon$ profiles around the entire pool would be necessary to determine the total TKE dissipation rate with better accuracy. However, this was not carried out due to time limitations, so it was assumed that the fitted red $\varepsilon$ profile shown in Fig. 5c was representative for the jet occurring in the gap area and the blue profile was representative for the turbulence occurring in the ice floe area. The areas were 1x3 $m^2$ on each side of the ice floe and 3x4 $m^2$, which respectively corresponds to the average gap between the floe and the pool boundaries and the floe itself (18 $m^2$ in total). The fitted profiles were integrated with respect to the depth and multiplied with their respective areas and with a water density of 1026 kg/$m^3$. The average total TKE dissipation rate was estimated to be 12.0 W, which means that approximately 60% of the mechanical input power to the system was dissipated in turbulence. However, the data scattering in $\varepsilon$ was quite large, approximately ±15% in E4 and as much as ±49% in E6 at the maximum values. Also note that this estimate is based on only one repetition at each ADCP position, hence should the result be used with caution.

The remaining 40% of the total input power must be dissipated by other mechanisms. Collisions between neighboring ice floes can for example lead to energy absorption during the impulse due to momentum transfer (Shen and Squire, 1998). This process may also occur during collision with an ice wall, which was the case in the current experiments. Energy was also transferred to surface waves in the pool and to water splash during the collisions, which was sometimes observed.

## CONCLUSIONS

Waves are effectively damped in fields covered by ice floes due to wave scattering, viscous dissipation inside the ice and energy dissipation caused by floe interactions. The dominating mechanism for wave energy dissipation in ice floe fields is not well known. In this work, the energy dissipation due to turbulence in the water around a moving ice floe has been investigated experimentally. Wave induced motion of a 3x4 m$^2$ ice floe has been simulated by pulling it back and forth in a 4x6 m$^2$ pool with 26 s oscillating periods. The axial motion and towing force applied on the ice floe was measured. This enabled calculation of average power input to the system, which was approximately 20 W.

Fluid motion in the water around the moving ice floe was measured with and ADV and an ADCP, which were installed next to the pool and in the floe center. The rate of TKE dissipation $\varepsilon$ was calculated from the estimated TKE spectra obtained from the two velocimeters. Total TKE dissipation rate was estimated to be 12 W by assuming uniform $\varepsilon$ profiles in the area covered by the ice floe and the gap between the floe and the pool, which was approximately 60% of the input power. However, there is a large uncertainty related to this number due to data scattering and few repetitions. Although the experimental setup was constructed (sharp edges and straight collision surfaces), the results indicate that a substantial portion of the wave energy is dissipated in turbulence when ice floes are set into motion which results in collisions.

## ACKNOWLEDGMENTS

The authors are grateful to Andrej Sliusarenko, Vladimir Markov and Sergej Podleshych for their assistance. We also thank Olav Gundersen for the winch setup. Funding for the experiment was provided by the Research Council of Norway under the PETROMAKS2 scheme (DOFI, Grant number 28062) and the IntPart project Arctic Offshore and Coastal Engineering in Changing Climate (Project number 274951). The data are available from the corresponding author upon request.